\documentclass{article}

\newcommand{\commenting}[1]{}

\usepackage[nonatbib, final]{ewrl_2024}

\usepackage[utf8]{inputenc} 
\usepackage[T1]{fontenc}    
\usepackage{hyperref}       
\usepackage{url}            
\usepackage{booktabs}       
\usepackage{amsfonts}       
\usepackage{nicefrac}       
\usepackage{microtype}      
\usepackage{xcolor}         
\usepackage{graphicx}
\usepackage{amsmath}
\usepackage{multirow}
\usepackage{lipsum}
\usepackage{bbm}
\usepackage{caption} 
\usepackage{subcaption}
\usepackage{float}

\title{Impact of Collective Behaviors of\\Autonomous Vehicles on Urban Traffic Dynamics:\\A Multi-Agent Reinforcement Learning Approach}
\author{
    Ahmet Onur Akman$^{1}$\thanks{First author}\\
    \texttt{onur.akman@uj.edu.pl}
    \And
    Anastasia Psarou$^{1}$\\
    \texttt{anastasia.psarou@uj.edu.pl}
    \And
    Zoltán György Varga$^{1}$\\
    \texttt{zoltan.varga@uj.edu.pl}
    \And
    Grzegorz Jamróz$^{2}$\\
    \texttt{grzegorz.jamroz@uj.edu.pl}
    \And
    Rafał Kucharski$^{2}$\\
    \texttt{rafal.kucharski@uj.edu.pl}
    \And
    \textnormal{$^1$ Doctoral School of Exact and Natural Sciences, Jagiellonian University, Kraków, Poland}\\
    \textnormal{$^2$ Faculty of Mathematics and Computer Science, Jagiellonian University, Kraków, Poland}
}

\commenting{
\author{
    \begin{tabular}{llllll}
        \multicolumn{2}{c}{Ahmet Onur Akman$^{1}$} & \multicolumn{2}{c}{Anastasia Psarou$^{1}$} & \multicolumn{2}{c}{Zoltán György Varga$^{1}$} \\
        \multicolumn{2}{c}{\texttt{onur.akman@uj.edu.pl}} & \multicolumn{2}{c}{\texttt{anastasia.psarou@uj.edu.pl}} & \multicolumn{2}{c}{\texttt{zoltan.varga@uj.edu.pl}} \\
        \\
        \multicolumn{3}{c}{Grzegorz Jamróz$^{2}$} & \multicolumn{3}{c}{Rafał Kucharski$^{2}$} \\
        \multicolumn{3}{c}{\texttt{grzegorz.jamroz@uj.edu.pl}} & \multicolumn{3}{c}{\texttt{rafal.kucharski@uj.edu.pl}} \\
        \\
        \multicolumn{6}{l}{$^1$ Doctoral School of Exact and Natural Sciences, Jagiellonian University, Kraków, Poland} \\
        \multicolumn{6}{l}{$^2$ Faculty of Mathematics and Computer Science, Jagiellonian University, Kraków, Poland}
    \end{tabular}
}
}

\begin{document}
\maketitle

\begin{abstract}

    This study examines the potential impact of reinforcement learning (RL)-enabled autonomous vehicles (AV) on urban traffic flow in a mixed traffic environment. We focus on a simplified day-to-day route choice problem in a multi-agent setting. We consider a city network where human drivers travel through their chosen routes to reach their destinations in minimum travel time. Then, we convert one-third of the population into AVs, which are RL agents employing Deep Q-learning algorithm. We define a set of optimization targets, or as we call them behaviors, namely selfish, collaborative, competitive, social, altruistic, and malicious. We impose a selected behavior on AVs through their rewards. We run our simulations using our in-house developed RL framework PARCOUR. Our simulations reveal that AVs optimize their travel times by up to 5\%, with varying impacts on human drivers' travel times depending on the AV behavior. In all cases where AVs adopt a self-serving behavior, they achieve shorter travel times than human drivers. Our findings highlight the complexity differences in learning tasks of each target behavior. We demonstrate that the multi-agent RL setting is applicable for collective routing on traffic networks, though their impact on coexisting parties greatly varies with the behaviors adopted.

\end{abstract}

\section{Introduction}
\label{sec:intro}

    
    With the recent advancements in autonomous driving technology, one can argue that we are moving towards a future where traffic systems are populated with self-driving autonomous vehicles (AV). Existing studies argue that the full integration will be costly and is still far down the road, but also suggest that AVs could significantly lower accident rates, increase accessibility, and improve overall traffic flow through optimized and coordinated route selection policies \cite{litman2020autonomous, martinez2018autonomous}. The current feasibility and anticipated prospects of integration of AVs are strongly promoted by parties with financial interests \cite{johnson2016peak, UKGovernment2025}.  However, managing this integration requires rigorous analysis of the multifaceted outcomes of such change, involving identifying and quantifying its impact on various stakeholders.
    
    One of the main affected parties in this transformation may be the current inhabitants of our traffic systems: human drivers. Apart from the aforementioned potential benefits, one of the reservations regarding autonomous driving technology is its potential to negatively impact the experience for human drivers, possibly exacerbating their travel times. Previous research has highlighted how this can occur naturally or consequently to a joint strategy \cite{metz2018developing, maciejewski2018congestion}. This research aims to contribute to these discussions by examining the interaction between human drivers and AVs within a simulated traffic environment. We demonstrate how different behavioral strategies of AVs can influence the traffic flow and how this can be simulated with multi-agent RL (MARL).
    
    
    In mixed traffic environments, the adopted behavior of a group of AVs may influence not only the efficiency of the group itself but also the overall traffic efficiency. Studies have shown that AVs can be programmed with various behavioral strategies and how different behaviors can affect traffic flow and congestion \cite{maciejewski2015large, biyik2020altruistic}. The interplay between these strategies and human driving behavior is an open field for investigation, as it determines to what extent AVs can enhance or disrupt existing traffic systems or compete with alternative modes to become the most attractive commuting option.
    
    Among the various challenges presented by the integration of AVs, we address the problem of route choice in mixed traffic environments. The route choice behaviors of the drivers can significantly influence the traffic flow. Human drivers typically base their route choices on experience, sometimes paired with real-time information, aiming to minimize their travel time \cite{ben1985discrete}. In contrast, AVs can employ more sophisticated learning algorithms and real-time data to make optimized route choices that consider both individual and collective benefit \cite{schwarting2018planning, li2015real}. This divergence in decision-making necessitates a detailed understanding of mixed route choice problems where humans and AVs coexist.
    
    
    In this study, we represent the discrete route choice problem within an RL framework by conceptualizing it as a multi-agent decision-making process. We construct a traffic environment, and we support the dynamics of this environment with traffic simulation software to reflect close-to-reality transport dynamics. In this traffic environment, a group of driver agents, consisting of human drivers and AVs, select routes to minimize their delays. We simulate the mobility of this driver population at a peak hour on a random day. The traffic system initially consists of only human drivers, modeled as selfish self-utility maximizers with no external guidance. After a predefined number of episodes, a portion of the human drivers is replaced by a group of AVs, which are RL agents employing Deep Q-Networks (DQN) \cite{mnih2013playing}. AVs learn a selected behavior imposed on them through their rewards. The reward formulation of an agent reflects the specific agent's behavior, such as minimizing one's own travel time or maximizing the delay for others. We investigate the outcomes in scenarios where AVs are imposed a selected behavior of each of six predefined behaviors: selfish, social, malicious, cooperative, competitive, and altruistic. We assess the attainability of these optimization targets for AVs in the given experimental setting and examine how they influence overall traffic efficiency.

    
    In the following, we provide more in-depth specifications for our problem formulation. Next, we introduce PARCOUR, our reinforcement learning framework that facilitated our experiments. We then present our findings, highlighting the interesting trends observed in different scenarios.  Our contributions can be summarized as follows:
    
    \begin{enumerate}
        \item Definition of a day-to-day route-choice problem in a traffic network and formulation of this problem as a MARL problem involving a varied set of agents.
        \item Development of PARCOUR, our reinforcement learning tool designed for custom learning scenarios like the one presented in this paper.
        \item Simulation of the proposed route-choice problem with different AV strategies and an analysis of our observations.
    \end{enumerate}
\section{Methodology}
\label{sec:method}

    \subsection{Scenario}
    \label{sec:scenario}

        We design a simplified transport system involving route choice and congestion. We simulate traffic movement at a peak hour on an arbitrary day. We provide each driver with a set of paths to choose from and a utility to maximize. We observe their experiences over the days and how they learn to optimize their returns.
        
        We create a population of 1200 human drivers intending to travel from their origins to their destinations through one of the three generated paths, by starting their commute on their individual start time. Each driver's start time is chosen randomly from the interval \((0, 3600)\), representing every second within an hour. The origin, destination, and start time information is initially assigned to each driver at random and remains unchanged throughout the simulation. Each human driver aims to reach their destination with minimal travel time. We allow this population to interact with the environment, learn about the traffic dynamics, and refine their knowledge about their options for 1000 episodes. We name this time period \emph{Phase Settle}.
        
        In \emph{Phase Shock}, starting in episode 1000, we replace approximately one-third of this population with AVs, modeling a sudden shift to autonomous driving in the city. We pause the learning process for humans and allow AVs to interact with this environment. By this, we aim to model the AVs' offline training procedure with a realistic environment model. Therefore, the remaining human drivers are still present in the environment with their frozen knowledge. We allow AVs to explore their options, and we once again allow the system to reach a loosely defined balance. 
        
        Finally in \emph{Phase Adapt}, which is from episode 4000, we resume the learning process for humans, while AVs are still learning. This is the time that AVs refine their offline-learned knowledge in the actual city traffic. Therefore, we now have two sets of agents in a shared system, adapting to each other's existence. We continue simulating for 2000 more episodes. We observe the implications of this coexistence for these two groups of drivers.

        \paragraph{Episodes} In each episode, we simulate driver movements with start times spread across an hour. Drivers learn from their route choices and the delays they experience upon these choices. They take turns to select routes, with the order determined by their start times. Before making their decisions, AVs receive information about routes chosen previously by others, which helps construct their agent state. Whereas human drivers decide solely based on their cost expectations associated with their options. After all route choices are made, we simulate the travel using traffic simulator software and collect each driver’s travel time. Agents then derive their rewards from this information according to their behavior definitions. For instance, every human driver is selfish; their rewards (to minimize) are their own travel times.

        \paragraph{Difference between AVs and humans} Humans are modeled using a behavioral model described in Section \ref{sec:human}, while AVs employ Deep Q-Learning from the RL framework. We define a set of reward-induced behaviors for AVs: selfish, collaborative, competitive, malicious, altruistic, and social. In each simulation run, we choose a different behavior to learn, but under the condition that the entire group of AVs will bear the same behavior. However, each of the behaviors are individual optimization targets and they do not constitute a collective objective, as we assume no direct communication or shared knowledge amongst our agents, including the AVs.

    
    \subsection{Traffic Network}
    \label{sec:network}

\begin{figure}[h!]
            \centering
            \begin{minipage}[b]{0.40\linewidth}
                \centering
                \includegraphics[width=\textwidth]{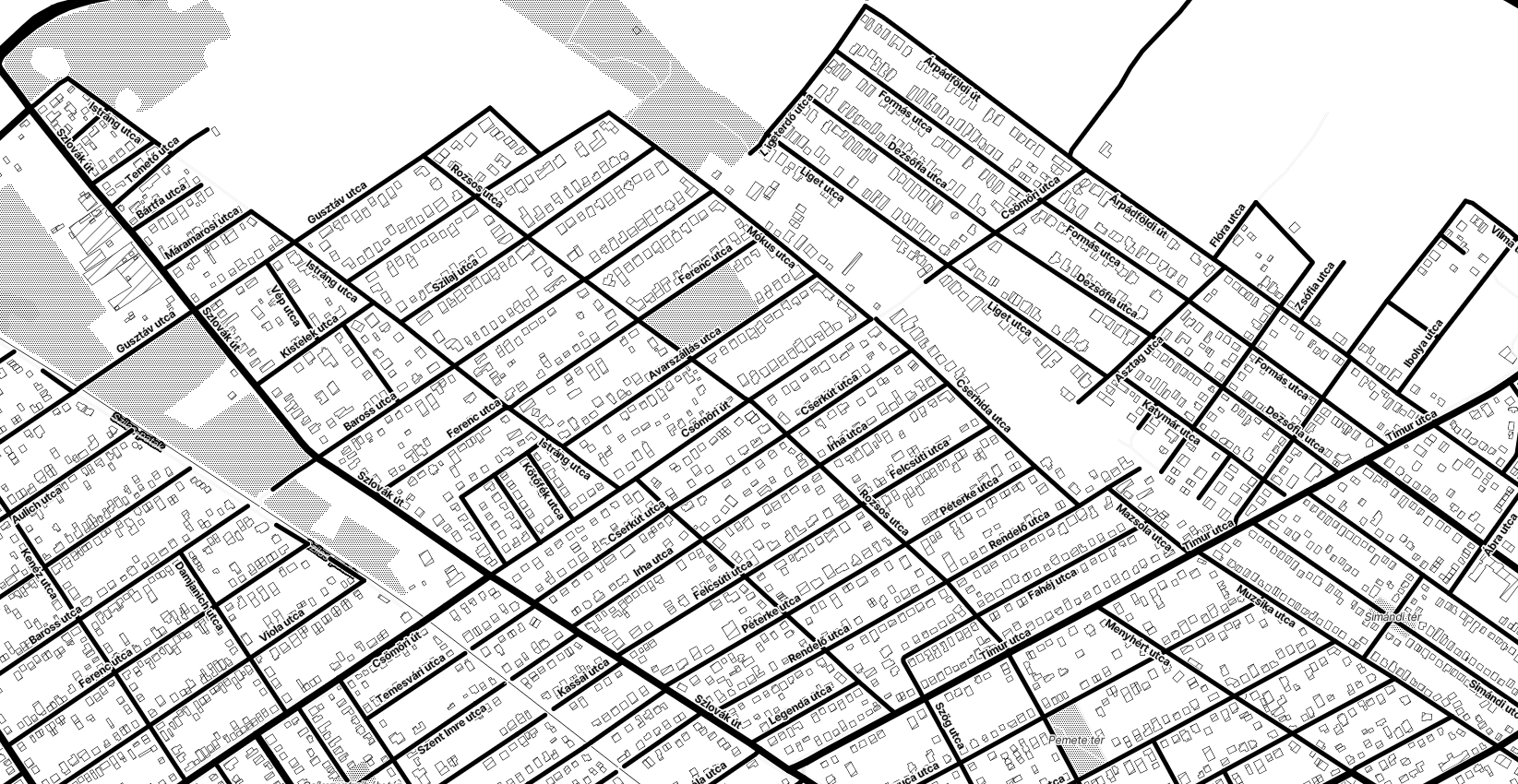}
                \caption{Csömör traffic network}
                \label{fig:csomor_newtork}
            \end{minipage}
            \hfill
            \begin{minipage}[b]{0.40\linewidth}
                \centering
                \includegraphics[width=\textwidth]{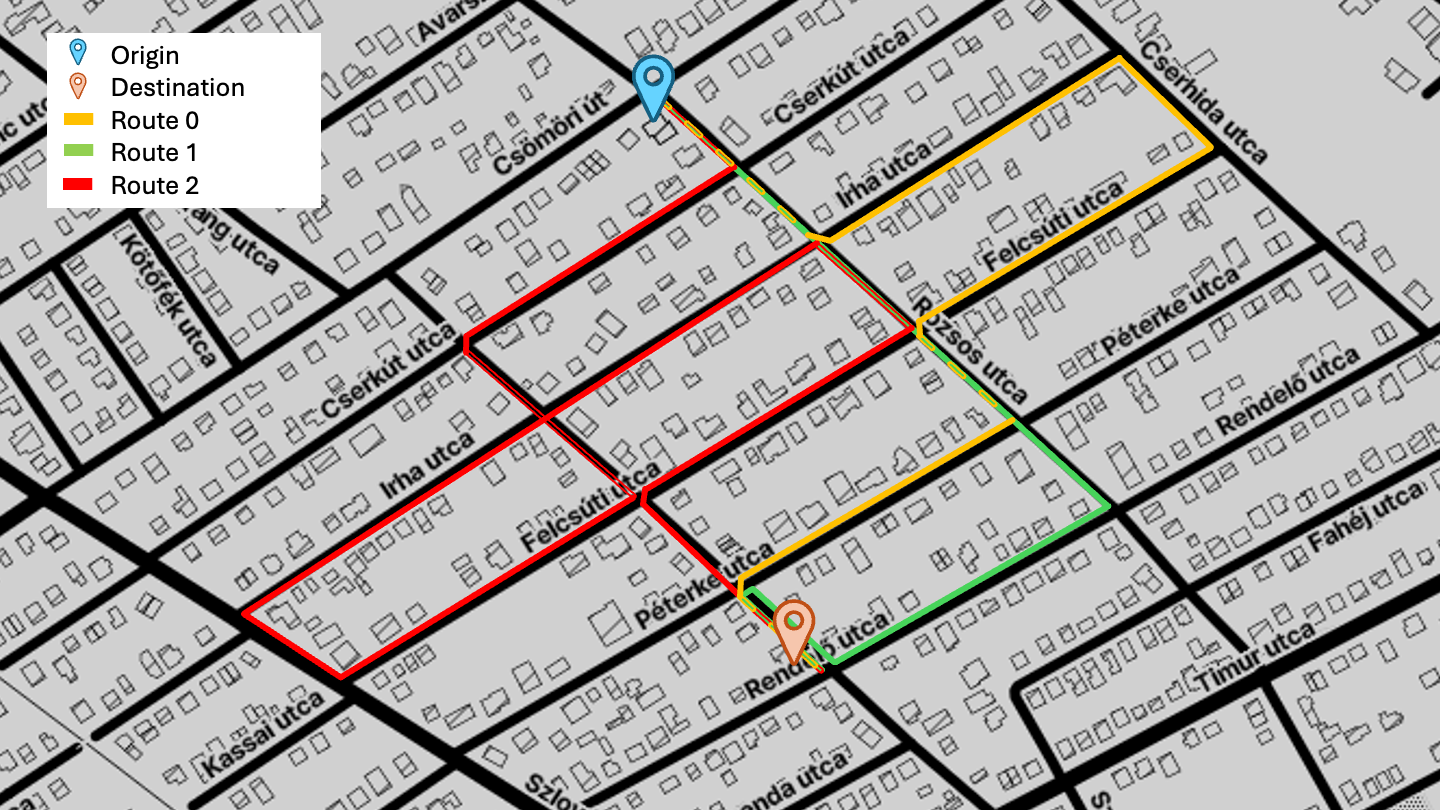}
                \caption{Generated paths for OD (0, 0)}
                \label{fig:routes_0_0}
            \end{minipage}
        \end{figure}

        We use the traffic network of Csömör, a small town in Hungary, shown in Fig. \ref{fig:csomor_newtork} \cite{StamenMaps}. This network is complex enough to demonstrate our case, with a well-organized grid-like layout with frequent junctions, facilitating necessary grounds for strategy development for our agents.  
        
\begin{figure}[h]
            \centering
            \includegraphics[width=0.55\linewidth]{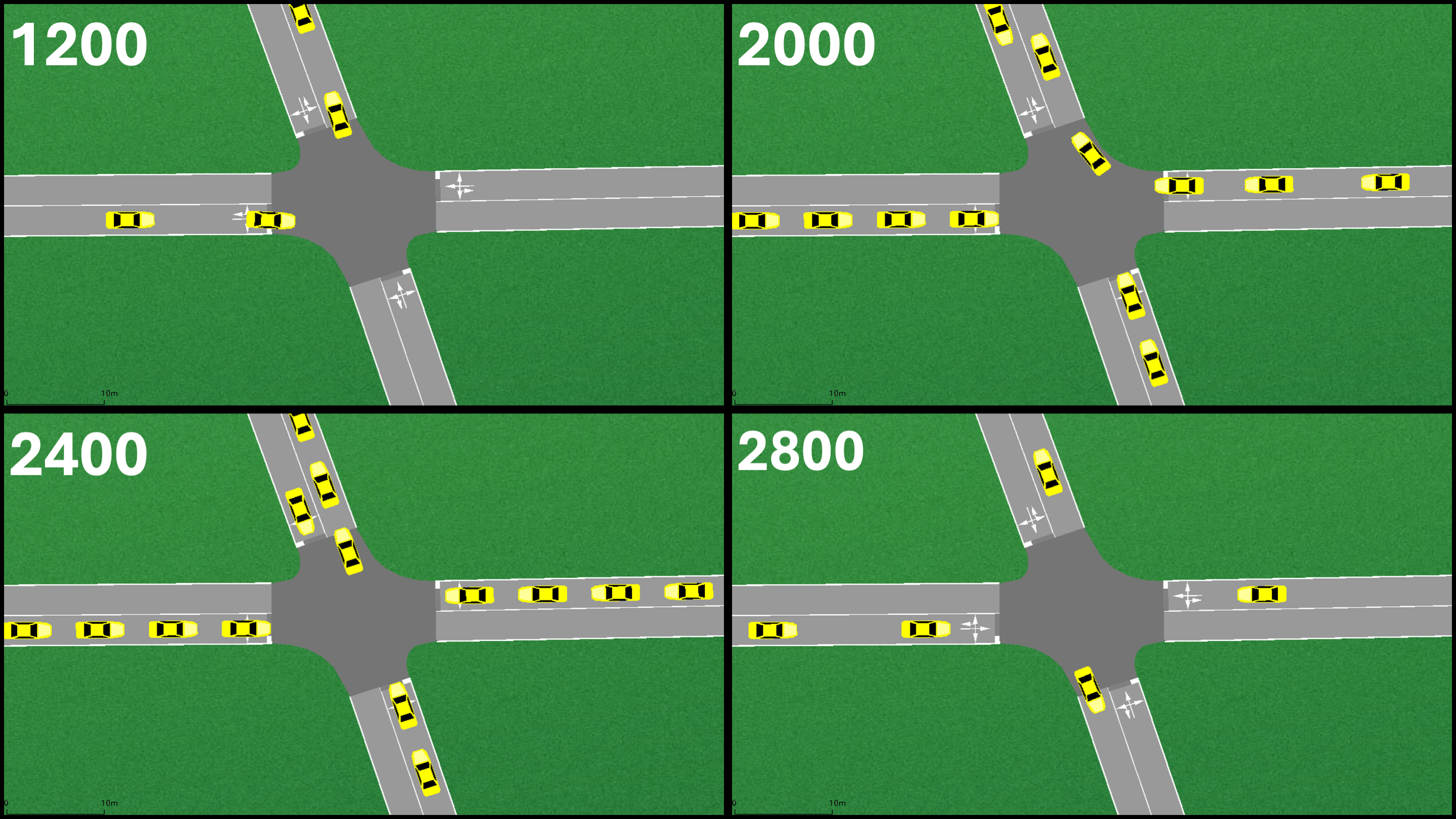}
            \caption{Congestion on a selected junction within a single episode on indicated time steps}
            \label{fig:congestions}
        \end{figure}

        We choose two origin and two destination nodes from this network and generate three paths connecting each of the four origin and destination combinations. We use a custom heuristic-based path-finding algorithm to create an action space that remains fixed across episodes for all agents. The resulting paths are not guaranteed to be optimal but sufficiently distinct, with a few intersections on critical nodes. We show all three paths generated to connect OD (0, 0) in Fig. \ref{fig:routes_0_0}.
        \paragraph{Congestion} We sample the agent start times from a Gaussian distribution centered on the half-hour mark within our one-hour range. This way, we observe congestion build up and incentivize the use of routes with higher free-flow travel time. The congestion on a selected junction over different time steps within the same episode is displayed in Figure \ref{fig:congestions}.


    \subsection{Human Modeling}
    \label{sec:human}

        We use the behavioral model presented in the literature \cite{cascetta2009transportation} for modeling our human agents' learning and decision-making procedure. This model is derived from random utility theory, a widely utilized theoretical paradigm for modeling transport-related choices among discrete alternatives. In each episode, the probability of an agent \(j\) to choose a route option \(i\) is:
        \begin{equation}
        \label{eq:logit}
        p_j(i) = 
        \frac
        {\exp{(\beta_j{\overline{c}_{i, j}})}}
        {\sum_{i'\in I^j} {\exp{(\beta_j{\overline{c}_{{i'},j}})}}} \,,
        \end{equation}
        where \(\overline{c}_{i, j}\) is agent \(j\)'s cost expectation for option \(i\). \(\beta_j\) represents the impact of human driver \(j\)'s personal traits on decision-making, which is assigned to each human agent from a predetermined negative range at random. This approach reflects our realistic assumption that not every human driver will have the same level of uncertainty involved in their decision-making.

        For each human agent \(j\), we define a memory of cost expectations for each available action. Each value in this memory is iteratively updated with perceived feedback as follows:
        \begin{equation}
        \label{eq:cost_learning}
        \overline{c}_{i, j}^{t+1} = 
        (1-\alpha_j) \cdot \overline{c}_{i, j}^t + 
        \alpha_j\cdot{c_{i,j}^t} \,,
        \end{equation}
        where \(\alpha_j\) is the fixed step size, \(c_{i,j}^t\) is agent \(j\)'s perception upon taking action \(i\) at time \(t\), \(\overline{c}_{i, j}^t\) and \(\overline{c}_{i, j}^{t+1}\) are the agent \(j\)'s expectations of costs associated with action \(i\) at time \(t\) and \(t+1\), respectively. In our case, \(c_{i,j}^t\) will be the time spent in route \(i\) from the start time until the arrival to the destination.

        In contrast to the RL algorithms, this model does not aim to represent an optimal decision-maker but rather to portray rational decision-making with some impact of personal traits and the human-like process of iterative expectation refinement. This model allows us to frame human driver behavior in a traditional agent-environment interaction loop as we have it for the RL agents.


    \subsection{Reward function and AV behaviors}
    \label{sec:rewards}

        The reward function imposes a selected behavior on the agent. For an agent \(k\) with behavioral parameters \(\varphi^k \in \mathbb{R}^4\), reward is obtained as:
        \begin{equation}
        \label{eq:reward}
        r_k = \varphi_1^k \cdot T_{own}^k + \varphi_2^k \cdot T_{group}^k + 
        \varphi_3^k \cdot T_{other}^k + \varphi_4^k \cdot T_{all}^k \,,
        \end{equation}
        where \(T^k\) is a vector of travel time statistics provided to agent \(k\), which contains:
        
        \begin{itemize}
            \item \textbf{Own travel time} \(T_{own}\): The amount of time the agent has spent in traffic.
            \item \textbf{Group's travel time} \(T_{group}\): The average travel time of agents within the same group as the given agent (e.g., AVs for an AV agent).
            \item \textbf{Other group's travel time} \(T_{other}\): The average travel time of agents within other groups than the given agent (e.g., humans for an AV agent).
            \item \textbf{System-wide travel time} \(T_{all}\): The average travel time of all the agents in the traffic.
        \end{itemize}

        We formulate each behavior with behavioral parameters listed in Table \ref{tab:behaviors}. We refer to behaviors with a positive \(\varphi_1\) as \emph{self-serving} behaviors. We care to maintain the same scale for the rewards across behaviors by choosing values with an absolute sum equal to \(1\). This allows us to observe comparable reward curves for different behaviors and maintain a consistent range of gradients in each experiment, providing comparable learning processes of deep networks.
        
        \begin{table}[h]
        \centering
        \caption{Behavioral strategies and their objective weightings}
            \begin{tabular}{c|cccc|c}
            \toprule
            \textbf{Behavior} & \textbf{\(\varphi_1\)} & \textbf{\(\varphi_2\)} & \textbf{\(\varphi_3\)} & \textbf{\(\varphi_4\)} & \textbf{Interpretation} \\ 
            \midrule
            Altruistic    & 0   & 0   & 0  & 1   & Minimize delay for everyone                             \\
            Collaborative & 0.5 & 0.5 & 0  & 0   & Minimize delay for oneself and one's own group          \\
            Competitive   & 2   & 0   & -1 & 0   & Minimize self-delay and maximise for other group        \\
            Malicious     & 0   & 0   & -1 & 0   & Maximise delay for other group                           \\
            Selfish       & 1   & 0   & 0  & 0   & Minimize delay for oneself                               \\
            Social        & 0.5 & 0   & 0  & 0.5 & Minimize delay for oneself and everyone                 \\ 
            \bottomrule
            \end{tabular}
        \label{tab:behaviors}
        \end{table}

        We assume that an agent's action can only impact a certain time window, and any events occurring outside of this window are not informative for the agent. Although the consequences of the agent’s decision might extend beyond this time window, we assume these effects are negligible. To avoid distracting factors, we calculate \(T^k\) for an agent \(k\) with start time \(t_k\) by considering only the traffic flow observed in the time window \([t_k-L_r, t_k+L_r]\), where \(L_r\) is a hyperparameter.

        
    \subsection{Observations and agent states}
    \label{sec:obs}

        We provide each agent with a partial statistic about the current traffic condition in the form of observations, and we derive it from other vehicles' decisions, leveraging the turn-based structure of our episodes, as detailed in Section \ref{sec:scenario}. The environment emits an observation \(o_t\) during each turn, which includes the selected actions in the previous turns within the same episode. We assume that an agent \(k\) with start time \(t_k\) can only observe a limited time window preceding \(t_k\). Therefore, we constrain the observations with an observation time window of a length \(L_o\), where \(L_o\) is a hyperparameter. We also limit the observations to include information only about those with the same OD pair as the observing agent.

        The observation \(o_t\) is used to derive the agent state \(s_{k, t}^a\) for agent \(k\). The agent state \(s_{t}^a\) embeds useful statistics about the current traffic conditions, leveraging the notion of action \emph{warmth} to maintain the temporal order of previous actions.
    
        Upon receiving the route choices of prior agents, agent \(k\) constructs its agent state \(s_{k, t}^a\) as a vector of size \(2 * |A|\). \(|A|=3\) is the number of route options for an agent and the same for all agents in our experiments. The first half of the values are filled with the \emph{warmth} of each action, but only considering prior agents from the same group. The second half is filled in the same way, but by considering the prior agents from other groups. The \emph{warmth} of an action \(j\) concerning the other group for an AV agent \(k\) with start time \(t_k\) is obtained as:
        \begin{equation}
        \label{eq:warmth}
            \textit{warmth}_j^{\text{other}} = 
            \sum_{i \in N_{\text{other}, w_{t_k}}} 
            (t_i - \min{w_{t_k}}) \cdot 
            \mathbbm{1}_{\{\text{OD}_i = \text{OD}_k\}} \cdot 
            \mathbbm{1}_{\{\text{action}_i = j\}} \,,
        \end{equation}
        where
        \begin{itemize}
            \item \(w_{t_k}\) is the observation time window of length \(L_o\) defined for start time \(t_k\). It contains all time steps ranging in \([t_k-L_o, t_k]\).
            \item \(N_{\text{other}, w_{t_k}}\) is the set of prior non-AV agents whose start times fall within the observation window \(w_{t_k}\).
            \item \(\text{OD}_i\) and \(\text{OD}_k\) are the OD pairs of agents \(i\) and \(k\), respectively.
            \item \(t_i\) and \(\text{action}_i\) are the start time and chosen action of agent \(i\), respectively.
            \item \(\mathbbm{1}_{\{\text{OD}_i = \text{OD}_k\}}\) is an indicator function that is \(1\) if the OD pair of agent \(i\) is the same as agent \(k\)'s, and 0 otherwise.
            \item \(\mathbbm{1}_{\{\text{action}_i = j\}}\) is an indicator function that is \(1\) if the action of the agent \(i\) is \(j\), and 0 otherwise.
        \end{itemize}
    
        Following this, an agent state \(s_{k, t}^a\) of AV agent \(k\) is constructed as:
        \begin{equation}
        \label{eq:state}
            s_{k, t}^a = 
            \begin{pmatrix}
            \textit{warmth}^{\text{AV}} \\ 
            \textit{warmth}^{\text{other}}
            \end{pmatrix}\,,
        \end{equation}
        where \(
        \textit{warmth}^{\text{AV}} = 
        \begin{pmatrix}
        \textit{warmth}_1^{\text{AV}} \\
        \textit{warmth}_2^{\text{AV}} \\
        \cdots \\
        \textit{warmth}_{|A|}^{\text{AV}} \\
        \end{pmatrix}
        \) and \(
        \textit{warmth}^{\text{other}} = 
        \begin{pmatrix}
        \textit{warmth}_1^{\text{other}} \\
        \textit{warmth}_2^{\text{other}} \\
        \cdots \\
        \textit{warmth}_{|A|}^{\text{other}} \\
        \end{pmatrix}
        \). In our presented scenario, for an AV agent, the set of other groups contains only the group of human drivers and vice versa.
    
        We should highlight that start times and OD pairs are intrinsically contained in the agent state. However, as these factors remain unmodified throughout the simulation runtime, we choose not to explicitly represent them. In other cases, one should modify the agent state accordingly.


    \subsection{PARCOUR}
    \label{sec:parcour}
    
        We conduct our experiments using our very own multi-agent reinforcement learning framework: PARCOUR. PARCOUR (Playground for Agents with Rationality Competing for Optimal Urban Routing) is designed to allow researchers to define and test different behavior and learning models in custom multi-agent route-choice scenarios and observe agent interactions in a shared environment.
        
        In PARCOUR, we ensure modularity by modeling each component with separate lifecycles and minimal dependencies. PARCOUR exemplifies the principle of separation of concerns (SoC), as each component has distinct responsibilities. Moreover, it relies solely on a small set of core libraries like PyTorch \cite{paszke2019pytorch} and pandas \cite{pandas} to provide broader compatibility. A Unified Modeling Language (UML) class diagram illustrating the structure of PARCOUR is provided in Appendix \ref{sec:uml_parcour}.

        PARCOUR is for discrete route choice problems, like the one we present in this study. A rigorous investigation of this type of problem requires an accurate modeling of the traffic flow. We capture the realistic transport dynamics by enabling the integration with an external traffic simulator. In our experiments, we choose to use Simulation of Urban MObility (SUMO) \cite{SUMO2018}, an open-source, microscopic, and continuous traffic simulation software \cite{behrisch2011sumo}. However, users can integrate any external traffic simulator with a Python interface by extending the provided \texttt{BaseSimulator} class. 
        
        The \texttt{ScenarioRunner} object serves as the central orchestrator, connecting and managing the interactions between various components. Users can model any scenario using their custom \texttt{ScenarioRunner} class. A scenario in PARCOUR is constructed as a series of \textit{phases}. The events to occur in each phase, the number of phases, and their spans are configurable by the user.

       The framework allows users to integrate an external RL library, or implement their own learning algorithms, just like the ones we utilize in this study. PARCOUR is still in early development and it is available as open-source software.\footnote{Available at: https://github.com/COeXISTENCE-PROJECT/parcour}
\section{Results}
\label{sec:results}

    We implement our scenario (Sec. \ref{sec:scenario}) in Csömör network (Sec. \ref{sec:network}) using PARCOUR, integrated with SUMO (Sec. \ref{sec:parcour}). We run experiments with six reward-induced AV behaviors (Sec. \ref{sec:rewards}) in identical settings. Each case was repeated three times, and the results were averaged to provide a clearer representation of the repeated trends and to capture the variability across trials. This section presents our findings and highlights the contrast between the impacts of different AV behaviors.


    \paragraph{Demand and Action Space} Our action space comprises 12 paths connecting every four OD pairs. Each route's free flow travel times are given in Table \ref{tab:ff_time}. Free flow travel time is the time it takes for a vehicle to complete its journey under no traffic congestion. The provided values are calculated by SUMO.

    \begin{table}[h!]
    \centering
    \caption{Free flow travel time (in minutes) for different routes}
        \begin{tabular}{c|c|ccc}
            \toprule
            \textbf{Origin} & \textbf{Destination} & \textbf{Route 0} & \textbf{Route 1} & \textbf{Route 2} \\
            \midrule
            0 & 0 & 2.08 & 1.39 & 3.22 \\ 
            0 & 1 & 0.78 & 2.95 & 2.36 \\ 
            1 & 0 & 0.69 & 0.92 & 1.88 \\ 
            1 & 1 & 1.91 & 3.02 & 1.99 \\ 
            \bottomrule
        \end{tabular}
    \label{tab:ff_time}
    \end{table}

\commenting{
    Each OD pair is associated with a demand load, measured by the number of drivers intending to travel between them at any timestep of the simulation. While we aim to maintain similar loads for each pair, perfect equality in demand is hardly the case in real-world traffic and is also not reflected in our generated driver population. Demand levels for each OD pair and AV to human driver ratios are shown in Figure \ref{fig:all_demand}.

    \begin{figure}[h!]
        \centering
        \includegraphics[width=0.3\linewidth]{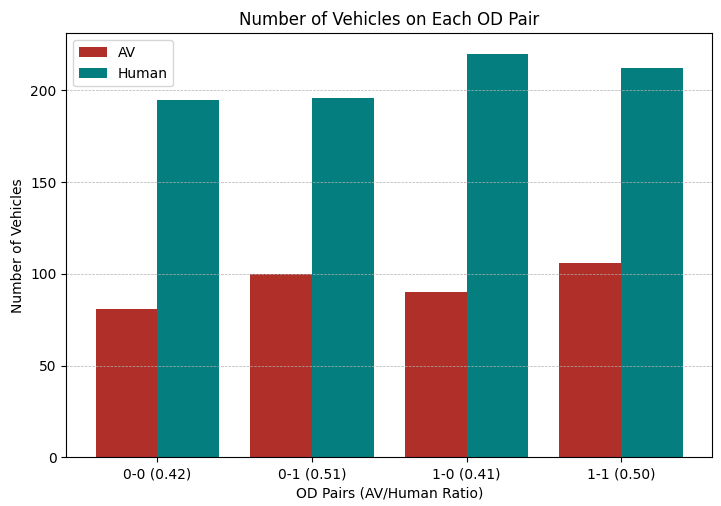}
        \caption{Number of drivers traveling on each OD. Up to Phase Shock, all demand is created by human drivers and the levels correspond to green and red bars combined.}
        \label{fig:all_demand}
    \end{figure}
}

    \noindent 
    \begin{minipage}[b]{0.45\textwidth}
        Each OD pair is associated with a demand load, measured by the number of drivers intending to travel between them at any timestep of the simulation. While we aim to maintain similar loads for each pair, perfect equality in demand is hardly the case in real-world traffic and is also not reflected in our generated driver population. Demand levels for each OD pair and AV to human driver ratios are shown in Figure \ref{fig:all_demand}. Note that up to Phase Shock, all demand is created by human drivers and the levels correspond to green and red bars combined.
    \end{minipage}%
    \hfill
    \begin{minipage}[b]{0.5\textwidth}
        \begin{figure}[H]
            \centering
            \includegraphics[width=0.85\linewidth]{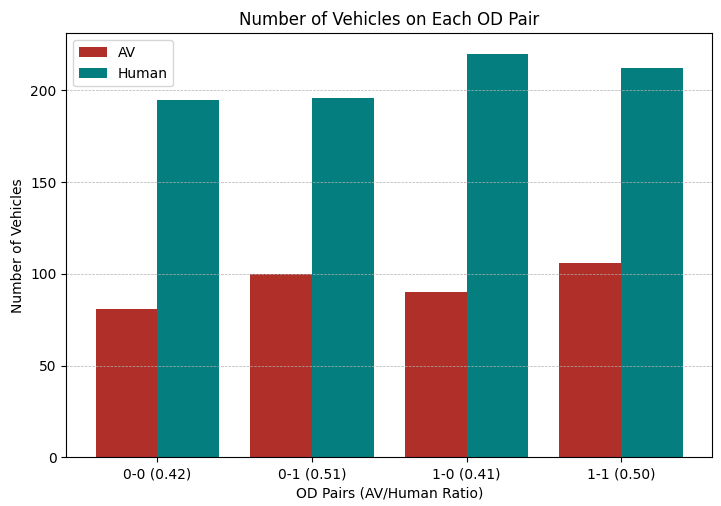}
            \caption{Number of drivers traveling on each OD}
            \label{fig:all_demand}
        \end{figure}
    \end{minipage}
    
\begin{figure}[h!]
        \centering
        \includegraphics[width=0.99\linewidth]{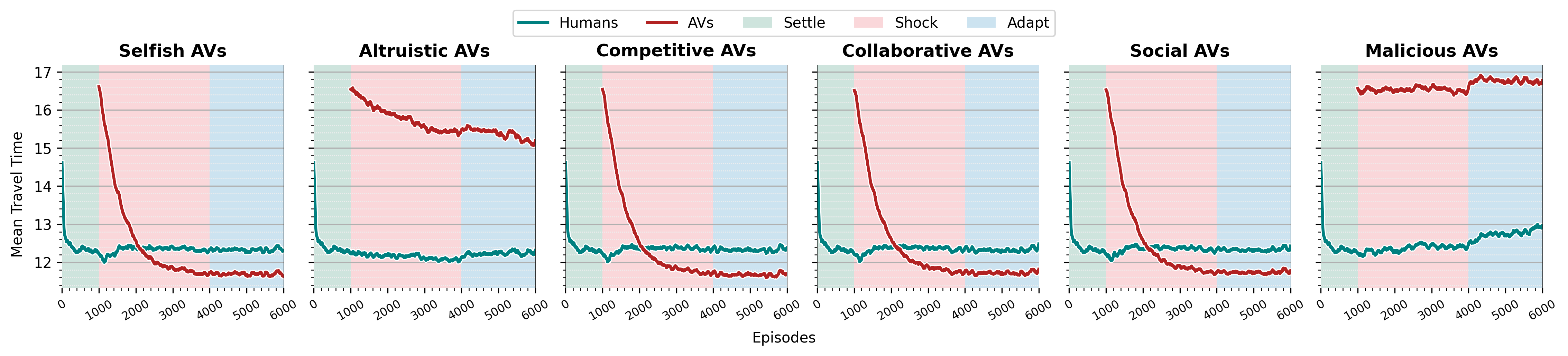}
        \caption{For each case, mean travel times of humans and AVs over episodes}
        \label{fig:all_tt}
    \end{figure}

    \paragraph{Comparing behaviors} We test six cases, in each imposing a different behavior on the AV fleet. We compare the mean travel times of humans and AVs in each experiment, with human driver data (origins, destinations, start times, if they will mutate to an AV) remaining consistent. This consistency yields similar curves in Phase Settle for humans. The mean travel times of AVs and humans over the episodes for each scenario are visualized in Figure \ref{fig:all_tt}.

    We show the consequences of different AV behaviors on their travel time and other drivers in Table \ref{tab:av_impact}, where we compare the mean travel times averaged on the last 100 episodes of Phase Settle and the last 100 episodes of Phase Adapt. Traffic efficiency measures the change in travel time of the entire driver population in the traffic system.

    \begin{table}[h!]
    \centering
    \caption{Impact of different AV behaviors. Positive impact indicates lowered travel time.}
        \begin{tabular}{l|ccc}
            \toprule
            \textbf{Behavior} & \textbf{On AV T.T.(\%)} & \textbf{On Human T.T. (\%)} & \textbf{On Traffic Efficiency (\%)} \\
            \midrule
            Altruistic    & -23.06 & 0.27  & -7.06 \\
            Collaborative & 4.28   & -0.71 & 0.86  \\
            Competitive   & 4.59   & -0.75  & 0.93  \\
            Malicious     & -36.29 & -5.43  & -15.12 \\
            Selfish       & 4.57  & -0.70  & 0.96  \\
            Social        & 4.21   & -0.66  & 0.87  \\
            \bottomrule
        \end{tabular}
        \label{tab:av_impact}
    \end{table}

    These results demonstrate the contrast in the impact of different AV behaviors on the traffic flow. The negative impact on the human commuting experience is relatively small but existent in competitive, collaborative, selfish, and social cases. All self-serving behaviors yield lower delays for AVs and increase overall traffic efficiency. In the malicious case, AVs establish a noticeable disadvantage for human drivers, at the cost of significantly more delays for themselves. Interestingly, malicious AVs manage to assert their malicious strategy more effectively once humans are allowed to react to their existence.  In the altruistic case, a minor enhancement in human driving experience costs \(23.06\%\) more delays for AVs. In our experimental setting, despite our definition of the altruistic objective, altruistic AVs not only fail to enhance traffic efficiency but worsen it.

\begin{figure}[h!]
        \centering
        \includegraphics[width=0.99\linewidth]{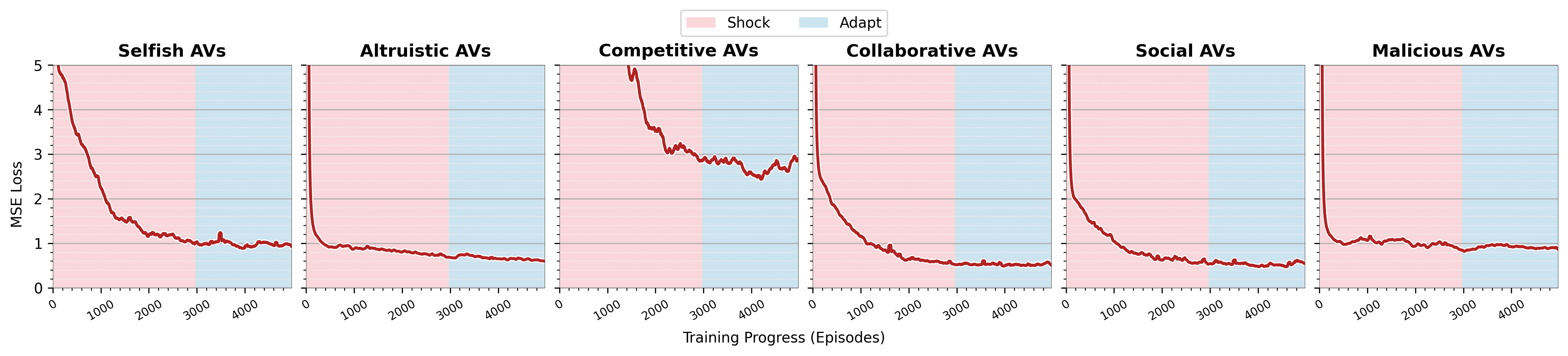}
        \caption{Average MSE loss of deep-RL agents over episodes}
        \label{fig:mse_loss}
    \end{figure}

    \paragraph{Learning efficiency of DQNs} We show the learning processes of DQNs of AVs in each case in Figure \ref{fig:mse_loss} using Mean Squared Error (MSE) loss. Our results show a generally downward trend in losses in all cases, highlighting an effective learning procedure in our simulations. Surprisingly, we observe that Phase Adapt is marked by further loss reduction and stabilization in most cases. However, depending on the AV objective, achieved losses and fluctuations vary greatly.
    
    Altruistic, collaborative, and social AVs exhibit a more consistent and effective learning process, converging steadily to a below-one MSE loss. In the malicious case, we see an early convergence to a near-one MSE loss from early on, but apart from a minor fluctuation at Phase Adapt, this level of loss is more or less preserved until the end of the simulation. Selfish AVs struggle relatively more to converge to an acceptable loss value, and we see that the loss curve is noisier, ending up with a below-one MSE loss. In contrast, competitive AVs exhibit a less stable learning curve, reflecting the increased complexity in the task of learning such a strategy in our multi-agent setting.

\begin{figure}[h!]
        \centering
        \includegraphics[width=0.99\linewidth]{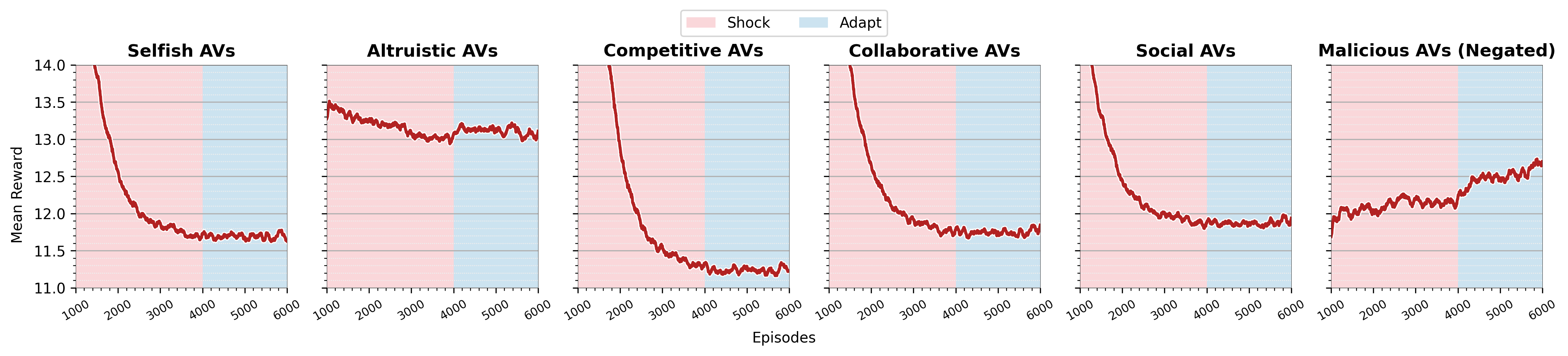}
        \caption{Average rewards of deep-RL agents over episodes observed in each experiment. Malicious rewards are negated for visualization.}
        \label{fig:rewards_all}
    \end{figure}

    To emphasize learning stability with reward acquisition, we show the reward curves for each case in Figure \ref{fig:rewards_all}. In the altruistic case, despite the stable loss curve, there is the least steady reward minimization over the episodes. This was anyway evident in Table \ref{tab:av_impact}. This hints at the difficulty of implementing such behavior in our experimental setting. We observe a similar situation in the malicious case, although Phase Adapt greatly aids the reward minimization (as also apparent in Figure \ref{fig:all_tt}). Other cases exhibit satisfactory reward minimization over the episodes. Interestingly, despite the unsteady loss curve, the competitive case exhibits the steepest improvement over the episodes, highlighting the contrast in difficulties in learning and implementing this behavior.

   \paragraph{Impact on human experience} Thus far we considered the changes in traffic dynamics as a whole and disregarded the differences between different parts of the traffic. However, to accurately assess the implementation and impact of each autonomous vehicle (AV) behavior, a more detailed, fine-grained analysis is required. This necessitates examining the differences across various subnetworks within the traffic system. Each set of routes connecting individual OD pairs constitutes a distinct subnetwork, and these subnetworks converge at key intersections. These subnetworks vary in terms of the number of nodes, path lengths, and the complexity of interactions at shared intersections, which are governed by priority-based turning rules. Therefore we expect to see variations in how AVs affect the driving experience in different parts of the traffic. 

\begin{figure}[h!]
        \centering
        \includegraphics[width=0.99\linewidth]{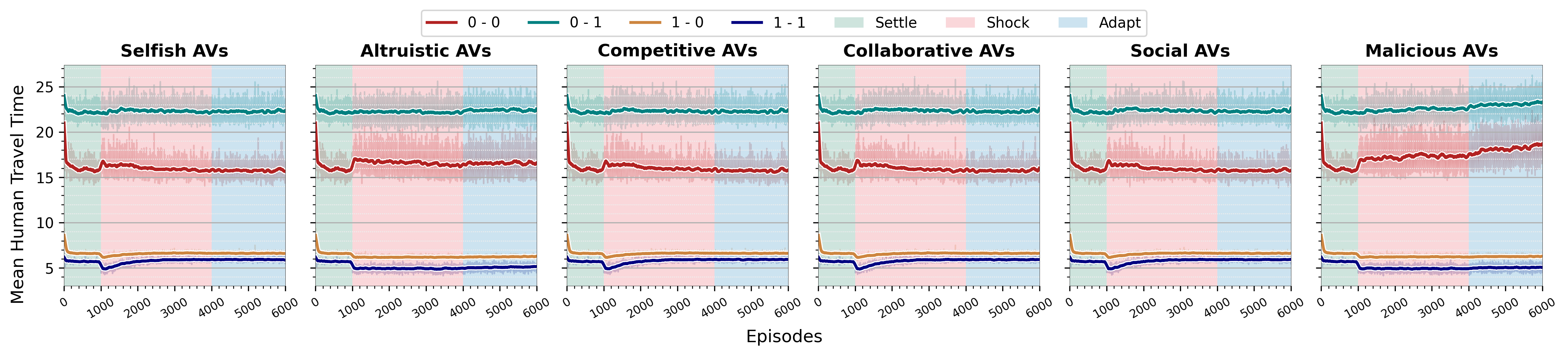}
        \caption{Change in human travel times for each OD pair. Error bars show the standard deviation across experiment runs.}
        \label{fig:human_tt_od}
    \end{figure}

   Firstly, we focus on how human driver travel times change over the episodes based on their origin and the destination they are traveling to. The shifts in human travel times for each OD pair are visualized in Figure \ref{fig:human_tt_od}. It is noticeable that the impact of AV behavior greatly varies across OD pairs. It is visible in all cases, and quite obvious in the malicious case, that human drivers in some subnetworks get affected negatively, while others either do not get affected or even experience a positive change in their travel times. We also notice that human travel times are more unpredictable in some ODs than others, indicated by wide error bars for ODs (0, 0) and (0, 1), while others are relatively consistent.

\begin{figure}[h!]
        \centering
        \includegraphics[width=0.99\linewidth]{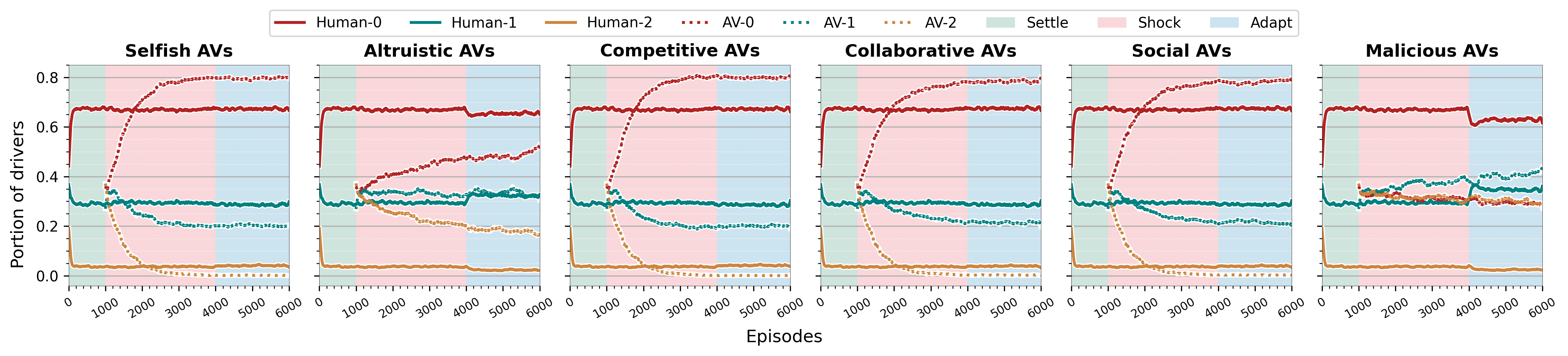}
        \caption{Portions of drivers choosing each route connecting origin 0 to destination 0}
        \label{fig:shift00}
    \end{figure}

    Next, we have a closer look at the changes that occurred in OD (0, 0) (which was displayed in Figure \ref{fig:routes_0_0}). In Figure \ref{fig:shift00}, we show the portion of each driver group choosing one of the three routes connecting OD (0, 0) over the episodes. This figure nicely visualizes the way AVs implement their strategies, and how humans adapt to them. The first thing we notice is that in all cases where AVs adopt a self-serving behavior, they optimize their travel time by simply populating the most favorable routes. We notice a slight increase in the number of human drivers in route 2, which is the least viable option. We interestingly observe quite similar changes in the cases of self-serving AVs and social AVs. In the altruistic case, we see that the AV domination in Route 0 is not as drastic, and AVs also populate less favorable routes as well. This makes Route 1 more attractive than how it was, and human drivers partially shift their preference from Routes 0 and 2 to Route 1.

    In the malicious case, AVs populate each route in a more balanced manner. After the Phase Adapt initiates, they favor Route 1 more, but the route preferences are still not as contrasted as in other cases. Interestingly, malicious AVs prefer Route 2 more than AVs in other cases, which is the longest path (as shown in Table \ref{tab:ff_time}), but we understand that this is for a good reason. Figure \ref{fig:routes_0_0} shows the critical points where three routes intersect. In a few spots, particularly at the intersection leading to the destination link, Route 2 has the priority. This may be consequential in congestion, like in the case at timestep 2400 in Figure \ref{fig:congestions}, as this may form long queues on the lanes with low priority. Given that malicious AVs focus solely on maximizing human travel times rather than minimizing their own delays, it becomes clear why selecting Route 2 can be a viable option for their strategy.

\begin{figure}[h!]
        \centering
        \includegraphics[width=0.99\linewidth]{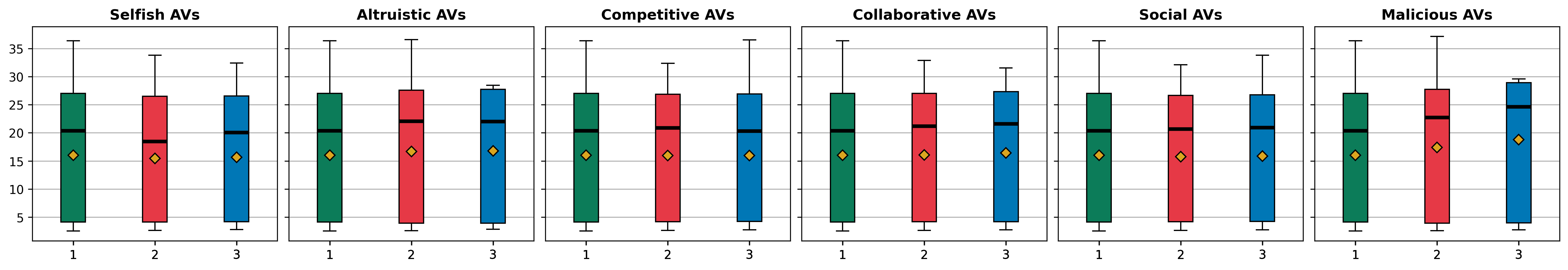}
        \caption{Distribution of travel times of human drivers traveling on OD (0, 0) in the final episode of (1) Phase Settle, (2) Phase Shock, (3) Phase Adapt. Black lines show the median and the golden markers show the mean points.}
        \label{fig:human_box}
    \end{figure}

    To understand how these preference changes affect travel times, we reveal the human travel time distributions in OD (0, 0) in Figure \ref{fig:human_box}. We observe noticeable changes in the spread of travel times, which indicates how AVs implement their strategies in Phase Shock and how humans react to these changes in Phase Adapt. For instance, we see that all self-serving behaviors cause tighter distributions of human travel times at the end of Phase Shock. Generally, we notice an increased spread for the cases where humans shift to Route 2, which is the least favorable route, and denser distributions for the cases where humans shift to Route 1, which is the middle ground in terms of attractiveness.

\section{Conclusions}
\label{sec:conclusions}

    We have conducted an experimental investigation of the potential implications of a shift to autonomous driving in transport systems. We defined a multi-agent route choice problem in a traffic network involving a driver population made up of humans and RL-enabled AVs. We established distinct learning and decision-making methods for both kinds, which enabled us to simulate their experiences in a unified RL agent-environment interaction loop.

    We introduced PARCOUR, our RL framework for route choice problems, which facilitated our experiments. PARCOUR enables the simulation of traffic flow in realistic transport dynamics by enabling integration with an external traffic simulator, which, in our case, was SUMO.

    We defined six target behaviors for AVs, each with a distinct impact on traffic flow. We formed a unified parameterization for formulating these behaviors through reward functions. We experimented with six scenarios where, in each case, the group of AVs adopted one of the defined behaviors. We analyzed the attainability of each of these behaviors in a MARL setting. We provided a clear overview of our findings using different data visualization techniques paired with our interpretations.
    
    Our results indicate that different reward definitions for AVs imply contrasted consequences for the other drivers in the traffic, as well as the users of the autonomous driving technology. We observed that each behavior can influence the overall traffic flow with different rates, while this is generally a positive influence for the self-serving AVs. We investigated the way AVs enforce their strategies, and we showed that the implications of their behaviors for different parts of the network vary greatly. We analyzed the preference changes of humans in adapting to the existence of AVs and showed that a portion of the human driver population needed to give up on their preferred routes.

    The multi-agent route choice problem concerning heterogeneous driver populations remains an open field of research. Drawing conclusions concerning real-world applications requires analyzing realistically scaled problems. With this study, we have provided a solid foundation for prospective studies, which will address more complex problem definitions, the use of more sophisticated learning models, the integration of a centralized control mechanism for AVs, and what other implications this may have on the human driving experience. 

    \paragraph{Author contributions} AOA - conceptualization, methodology, development, experimentation, data analysis, visualization, writing, review, editing; AP - methodology, development; ZGV - methodology, development; GJ - conceptualization, methodology, review; RK - conceptualization, methodology, review, project administration, funding acquisition.

    \paragraph{Acknowledgement} This research is financed by the European Union within the Horizon Europe Framework Programme, ERC Starting Grant number 101075838: COeXISTENCE.

\newpage
\bibliographystyle{plain}
\bibliography{ewrl_2024}

\newpage
\appendix

\section{UML class diagram of PARCOUR}
\label{sec:uml_parcour}

\begin{figure}[h!]
        \centering
        \includegraphics[width=0.9\linewidth]{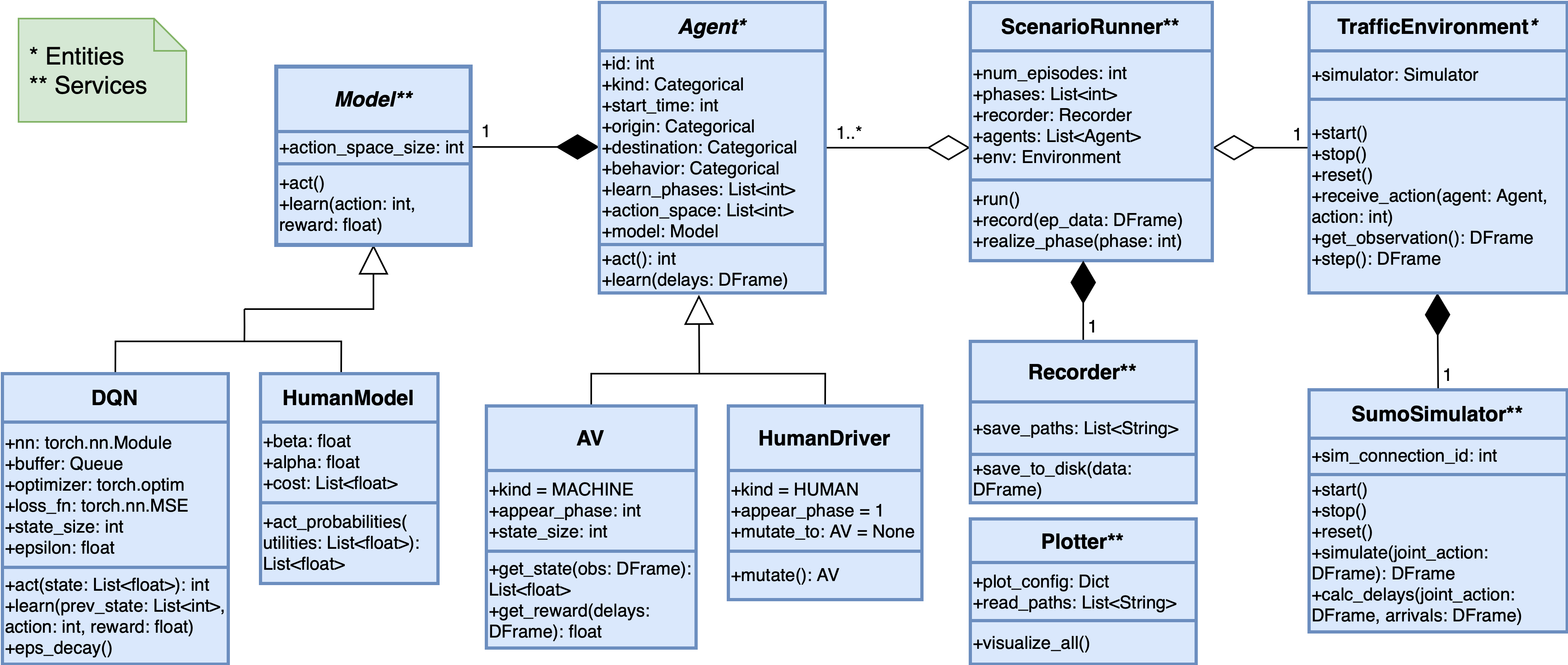}
        \caption{UML Class Diagram of PARCOUR}
        \label{fig:parcour_uml}
    \end{figure}

    Figure \ref{fig:parcour_uml} displays a UML class diagram illustrating the structure of PARCOUR. This diagram presents only the classes we used in our experiments.

\section{Reproducibility}
\label{sec:repro}

    The results we present in this paper are obtained through experiments with \(1200\) driver agents and \(6000\) episodes. Generated driver population data is provided in the same repository as PARCOUR. We construct the scenario we describe in Section \ref{sec:scenario} with 3 phases, namely: Settle, Shock, and Adapt. These phases start in episodes \(1\), \(1000\), and \(4000\), respectively. We conducted each experiment three times, and the results presented are the averages of these repetitions.

    Phase Settle is where all driver agents are humans, implemented using the model we describe in Section \ref{sec:human}. Phase Shock is where \(377\) randomly selected human drivers are replaced by AVs. AVs are RL agents utilizing the (Single) Deep Q-learning algorithm. We turn off the learning for humans in Phase Shock. Phase Adapt is when we enable learning for everyone and let both parties adjust to each other's existence.

    Start time and OD assignments to agents are done randomly. OD assignment distributions are provided in Figure \ref{fig:all_demand}. Start times range in a period of one hour, with seconds precision. We sample start times from a Gaussian distribution as justified in Section \ref{sec:network}.

    AVs all have their individual Q-Networks, trained solely using their own experiences. Each network has an identical architecture, implemented using \texttt{nn.Module} from PyTorch. Each network consists of an input layer, two hidden layers, and an output layer. It processes the state input through the input layer to \(32\) units, followed by a hidden layer to \(64\) units and another hidden layer to \(32\) units before mapping values to the action through the output layer. ReLU activation is applied after each layer except the output. The learning rate is the same for all agents, fixed to \(0.003\). We use MSE loss and Adam optimizer, which are implemented in \texttt{PyTorch}. Each agent has a replay buffer of size \(256\) and a batch size \(32\). The exploration is ensured by an epsilon-greedy policy with \(\varepsilon\) set to \(0.99\) and a decay rate of \(0.998\), applied in every episode.
    
    The observation window size \(L_o\) (Sec. \ref{sec:obs}) is set to \(300\), which corresponds to a \(5\) minutes range. The reward window size \(L_r\) (Sec. \ref{sec:rewards}) is set equal to \(L_o\).

    Humans have their individual cost tables (Sec. \ref{sec:human}), which we initially populated with each route option's free flow travel times. They share the same \(\alpha\) value, set to \(0.2\). They have randomized \(\beta\), as reasoned in Section \ref{sec:human}, randomly determined from the range \([-0.8, -0.2]\).

    The network we used is included within the same repository as PARCOUR. We use nodes from our traffic network \texttt{279952229\#0} and \texttt{115604053} as origins, and \texttt{-115602933\#2} and \texttt{-441496282\#1} as destinations (Sec. \ref{sec:network}). We generate the action space using our path generation algorithm with logit \(\beta\) set to \(-0.1\), creating three routes linking every origin-destination combination. Generated paths and the path generation code are also included in our repository.

    PARCOUR has an integrated plotting functionality, which automatically saves useful plots to the disk once an experiment concludes. However, the plots we present in this work are produced using a different code from the data generated in our experiments. All line plots in Section \ref{sec:results} are smoothed by \(50\) steps using \texttt{uniform\_filter1d} from \texttt{scipy}.
    

\section{Computing Resources}
\label{sec:comp_resources}

The experiments (Sec. \ref{sec:results}) were conducted on our faculty's high-performance computing cluster. The resources allocated for these experiments included a single GPU (NVIDIA Tesla V100), 64 GB of RAM, and 4 CPU cores per task. The total execution time for each experiment was around 18 hours.


\end{document}